\documentclass[12pt]{article}
\usepackage{epsfig}
\usepackage{cite}

\begin{document}

\title{Length Variations of European Baselines Derived from VLBI and GPS Observations\footnote%
  {In: D. Behrend, A. Rius (Eds.), Proc. 15th Working Meeting on European VLBI for Geodesy and Astrometry, Barcelona, Spain, Sep 07-08, 2001, 116-123.}}
\author{Zinovy~Malkin, Natalia~Panafidina, Elena~Skurikhina}
\date{\vspace{-10mm}}
\maketitle

\begin{abstract}
Results of VLBI and GPS observations were analyzed with goal
to investigate differences in observed baseline length derived
from both techniques.  VLBI coordinates for European stations
were obtained from processing of all available observations
collected on European and global VLBI network.  Advanced model
for antenna thermal deformation was applied to account for
change of horizontal component of baseline length.  GPS data
were obtained from re-processing of the weekly EPN (European Permanent
GPS Network) solutions.
Systematic differences between results obtained with two
techniques including linear drift and seasonal effects are determined.
\end{abstract}

\section{Introduction}

European region is one of the most intensively studied areas of the Earth
from the point of view of regional geodynamics. There are more than 100
permanently operating GPS receivers, about 10 permanent VLBI stations
and more than 10 permanent SLR stations. Lately
much attention has been devoted to comparison and combination of results
obtained using different space geodesy techniques.  This work is devoted
to comparison of baseline length variations derived from GPS and VLBI observations,
continuing the cycle of works on this
problem, see e.g. \cite{Campbell00a,Campbell00b,Haas00,Lanotte99,%
Malkin01h,GSFC00,Titov99,Tornatore99}.
It should be mentioned here that observed changes of the baseline length on the
one hand are resulted by insufficient corrections for
observational effects such as thermal deformations of
VLBI antennas, for example, or errors in modeling of tropospheric
refraction, but on the other hand they are subjected to a number of
insufficiently studied or not taken into account properly geophysical effects
that can result in the real changes of the baseline length, these effects may be
atmospheric and snow loading, tides, postglacial rebound, and so on.
It is important that the majority of these effects has both seasonal
and secular components.

In this study we have analyzed VLBI and GPS observations at 6 European
stations carrying out both VLBI and GPS regular observations and having long
enough observational history.

\section{Data used}

\subsection{VLBI observations}

VLBI baseline lengths were computed with the OCCAM package using all available 24h
sessions for the period of 1983.9--2001.5.
Details of the method used can be found in \cite{Skurikhina00}.

Linear trend in the baseline lengths was computed for the whole period of
observations and, for more accurate comparison with GPS data, for the
period of 1996.0-2001.0.  Only later results are presented in this study.
In \cite{Malkin01h} we compared linear trend in variations of
baseline length derived from the observations over the period 1996.0--2001.0
with ones computed using all available VLBI sessions over a period
1990.0--2001.4.
Differences of estimated rates are inside one sigma interval for all
baselines analyzed here.

For more strict account for variation in baseline lengths due to thermal
antenna deformations we used advanced model of this effect
\cite{Skurikhina01} which allow to correct observed station position
not only for vertical but also for horizontal displacement.
At this stage of the research we used zero time delay between
change of air and telescope construction temperature, because of
lack of such a data for most of antennas. However, this mismodelling will
effect only intra-day displacement of the telesope reference point,
but not seasonal variations.

It should be mentioned here that account for horizontal displacement
is especially important for processing regional networks, whereas
vertical displacement due to thermal deformations prevails in
variations of global baselines length. In particular, errors in modelling
of this effect may be a possible reason of seasonal baseline length variations
found e.g. in \cite{Titov00}.

\begin{figure}
\centering
\epsfxsize=\textwidth \epsffile{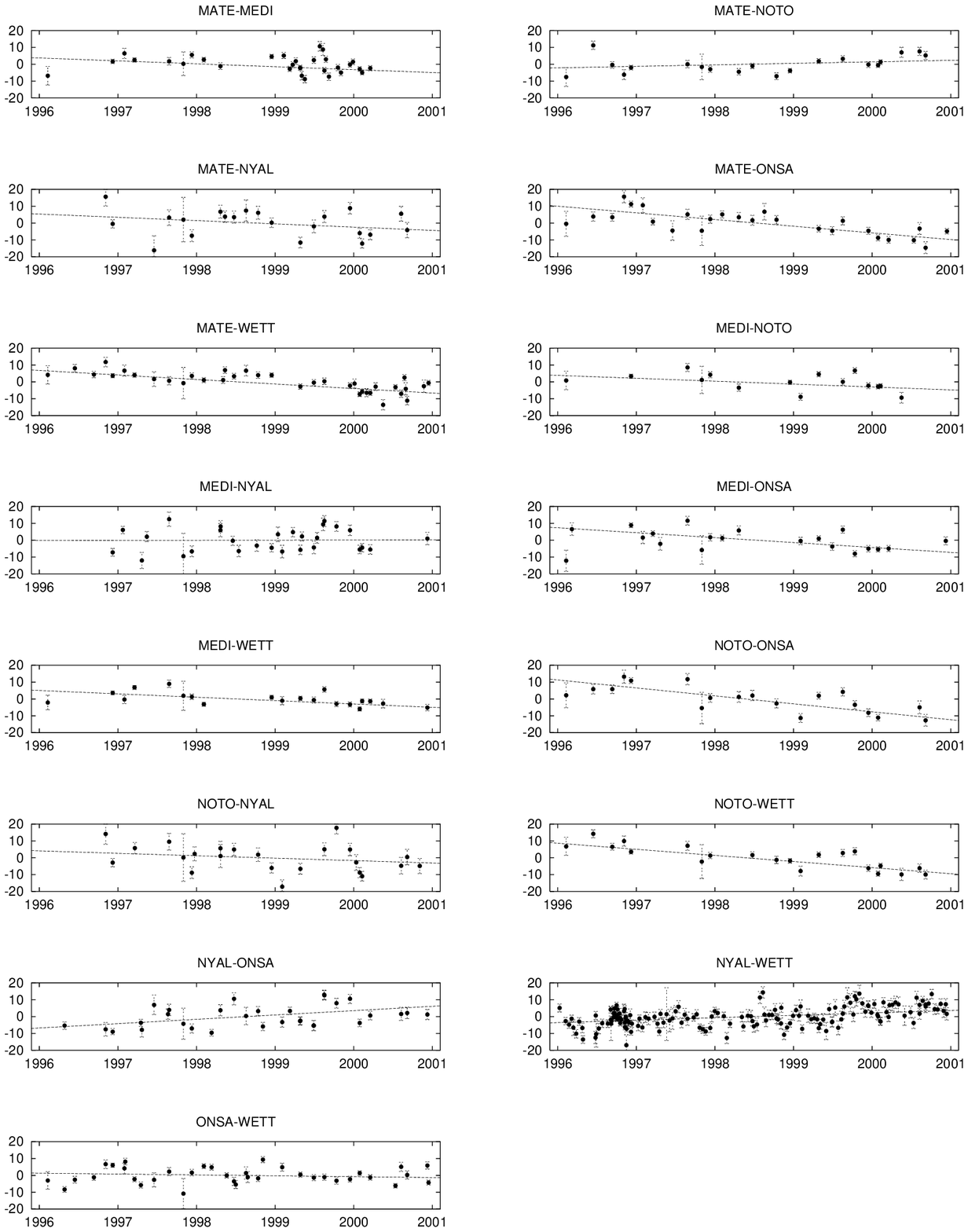}
\caption{Variation of VLBI baseline lengths, mm.}
\label{fig:vlbi}
\end{figure}

Variation of baseline lengths obtained from VLBI data are shown in
Figure~\ref{fig:vlbi}.
Unfortunately, stations Crimea and Yebes are not equipped with GPS
receiver.

\subsection{GPS observations}

For computation of baseline lengths between European GPS stations
we used weekly EPN solutions distributed in SINEX files.  However,
this solutions are not suitable for immediate use in geodynamical
analysis because they cannot provide homogeneous long-time coordinate
time series due to periodic changes in reference coordinate system
and set of fiducial stations.
For this reason, direct use of the EUREF solutions shows jumps in
baseline length variations \cite{Lanotte99}.  Besides, method of
computation of station coordinates used in EPN is based on using
tight constrains to fiducial stations which cause a distortion
of the network, i.e. fictive variations in baseline lengths
(see e.g. \cite{Malkin01g}).

So, variations of baseline lengths from GPS data were obtained from analysis
of coordinate time series for EPN stations computed by the method
described in \cite{Malkin01j}.  This computation is based on
de-constraining of the official EPN solutions with further transformation
to ITRF2000.  For this study we used 6-parameter Helmert transformation
to avoid loss of seasonal geophysical signal in baseline length.
Using our independent
coordinate time series allows us to obtain realistic station displacement
practically free of network distortion
for all EPN stations over the period 1996.0--2001.0.

\begin{figure}
\centering
\epsfxsize=\textwidth \epsffile{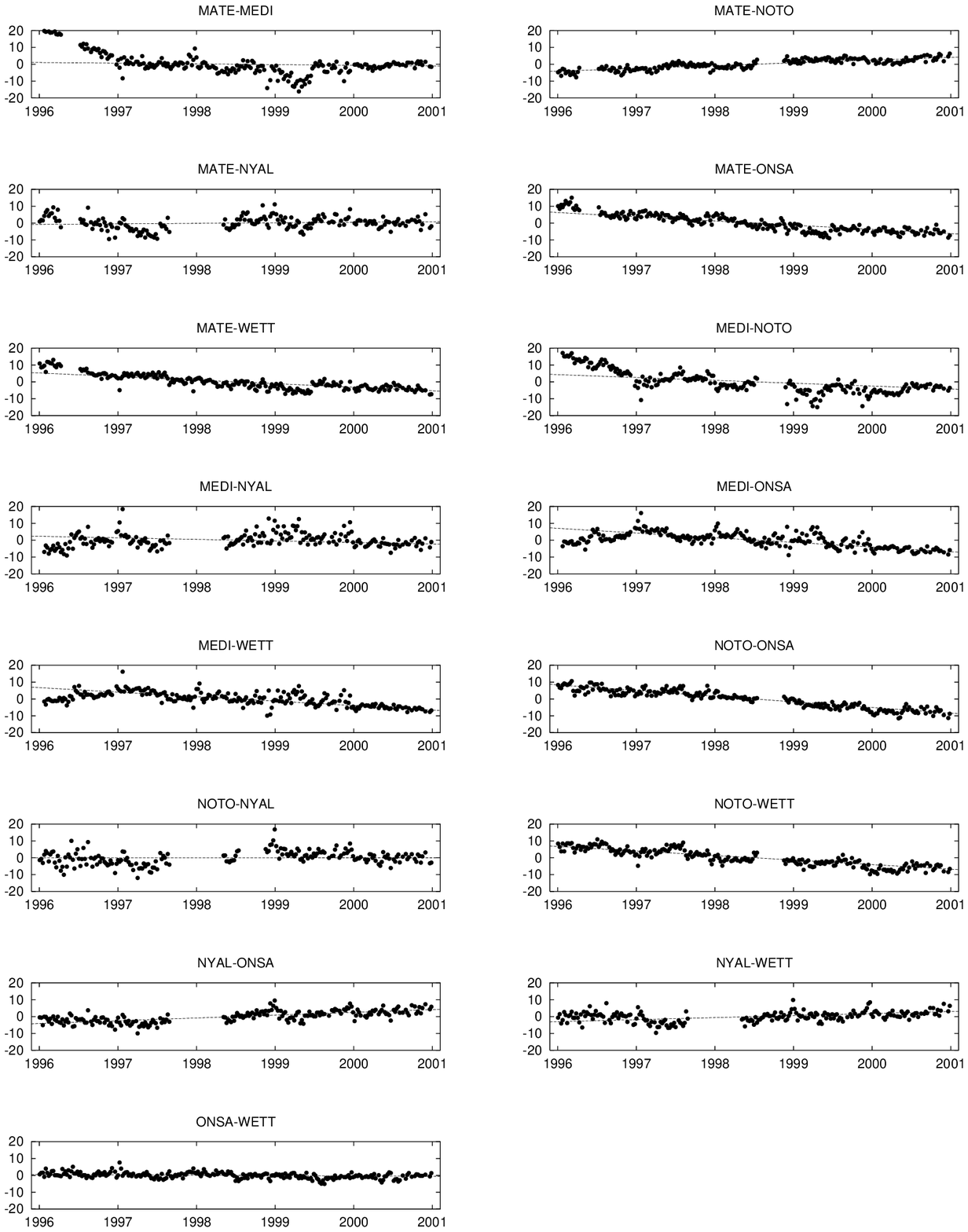}
\caption{Variation of GPS baseline lengths, mm.}
\label{fig:gps}
\end{figure}

Variation of baseline lengths obtained from GPS data are shown in
Figure~\ref{fig:gps}.
Unfortunately, MADR coordinate time series is too short (less
than two years) that does not allow to get reliable results.
It should be mentioned that errors in GPS baseline length significantly decrease
during a period under investigation.
E.g., one can see abnormal trend in MATE displacement in 1996.
However, that does not influence result very much due to relatively
small weight of these data.

\subsection{Atmospheric loading}

\begin{figure}
\centering
\epsfxsize=\textwidth \epsffile{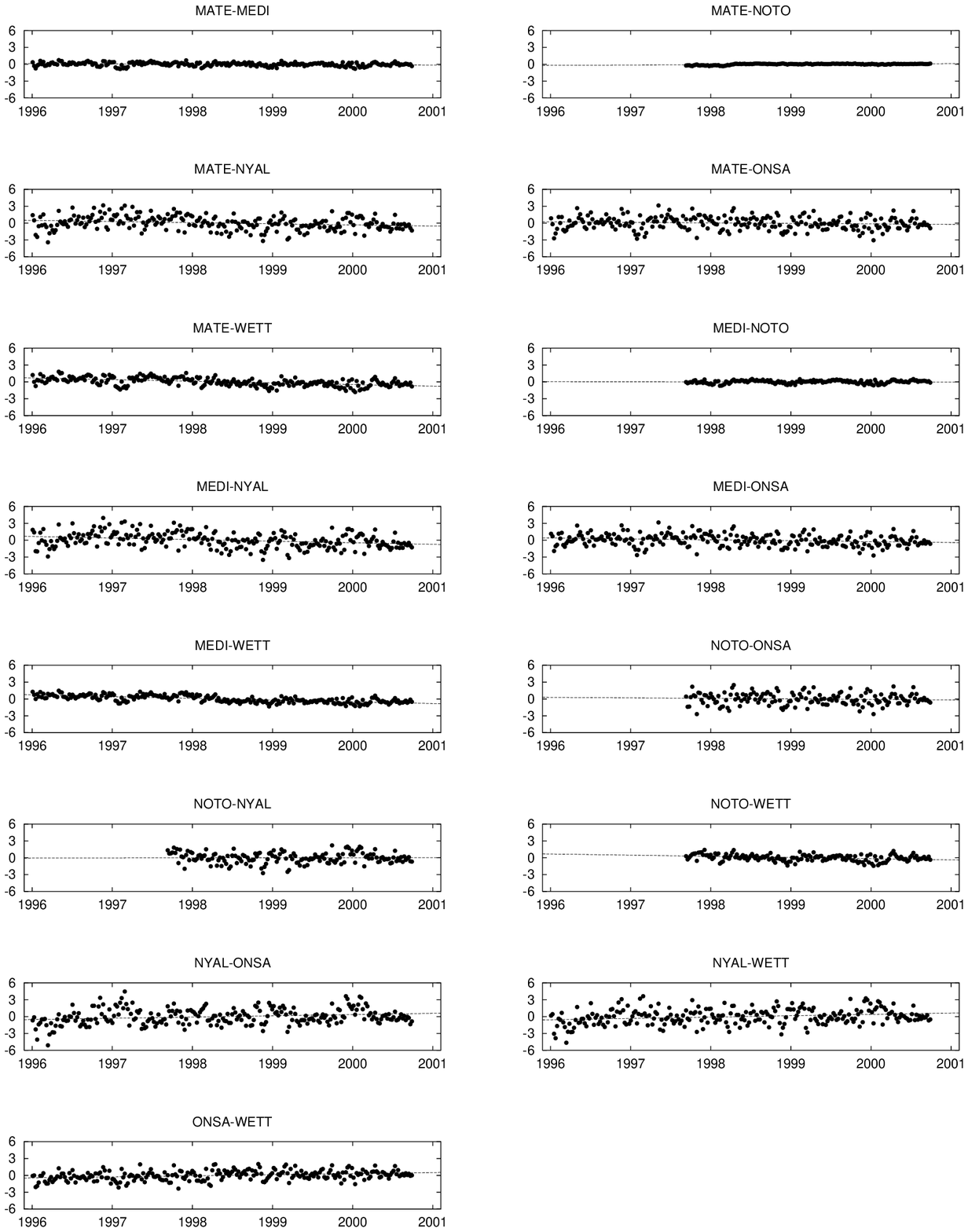}
\caption{Variation of baseline lengths due to atmospheric loading, mm.}
\label{fig:atmosph}
\end{figure}

One of the most important factor affecting variations of station coordinates
derived from space geodesy observations is atmospheric loading.
We investigated influence of this effect using 3-dimensional atmospheric
loading time series provided by H.-G.~Scherneck \cite{Atmloading}.
The data were averaged over a week interval corresponding to every GPS week
and variations of baseline lengths were computed from these weekly values.
Variation of baseline lengths obtained from analysis of atmospheric data
are shown in Figure~\ref{fig:atmosph}.

It is interesting that baseline length
variations contain not only seasonal but also secular component, even for
short baselines, especially for continental-coastal ones, in particular
baselines including Wettzell stations which are often used in studies
on European geodynamics, e.g. \cite{Campbell00b,Haas00,Lanotte99,Tornatore99}.
The reason of that may be long-periodic or progressive weather and
climate changes, but period of our investigation is too short to separate
them.

Since variations in height component of station displacement due to
atmospheric loading prevail, this effect is especially significant
for global baselines.  For regional networks horizontal displacements
yield main contribution to variation of baseline lengths.

\section{Results and conclusions}

Results of computation of variations in baseline lengths are
presented in \ref{tab:result}. One can see that values of rates
obtained from VLBI and GPS observations are in good agreement
for most baselines. Unfortunately, it is not the case for seasonal
variations. Obviously, interval of investigation is too short and
number of used VLBI observations is too small for many baselines.

Indeed, it would be important to verify our results using data obtained from
other space geodesy techniques, but only MATE and WETT stations are
equipped with SLR units, and only NYAL station is equipped with DORIS
beacon (which is explained by difficulties in collocation of DORIS beacon
and VLBI antenna due to radio frequency interference).

\begin{table} 
\centering
\caption{Results of analysis of variation of baseline lengths: baseline length (L),
km, number of epochs (N) processed and found in the IVS data base,
linear trend (Rate), mm, amplitude of annual term (As),
mm, amplitude of semiannual term (Asa), mm.}
\label{tab:result}
\tabcolsep=4pt           
\bigskip
\begin{tabular}{|c|c|crrr|crrr|rrr|}
\hline
Base & L & \multicolumn{4}{|c|}{VLBI} & \multicolumn{4}{|c|}{GPS} & \multicolumn{3}{|c|}{Atmospheric loading} \\
\cline{3-13}
&& N & Rate & Aa & Asa & N & Rate & Aa & Asa & Rate & Aa & Asa \\
\hline
\hline
MATE &  597 &~31 & --1.78    &  0.37     &  2.23     & 224 & --0.97    &  3.42     &  1.26     & --0.05    &  0.16     &  0.05     \\
MEDI &      &~35 & $\pm0.63$ & $\pm0.99$ & $\pm1.12$ &     & $\pm0.21$ & $\pm0.43$ & $\pm0.36$ & $\pm0.01$ & $\pm0.03$ & $\pm0.03$ \\
\hline
MATE & ~444 &~20 & +0.62     &  3.21     &  2.17     & 212 & +1.57     &  0.51     &  0.35     & +0.06     &  0.06     &  0.02     \\
NOTO &      &~23 & $\pm0.71$ & $\pm1.16$ & $\pm1.29$ &     & $\pm0.10$ & $\pm0.19$ & $\pm0.16$ & $\pm0.01$ & $\pm0.01$ & $\pm0.01$ \\
\hline
MATE & 4190 &~21 & +0.27     &  6.57     &  3.35     & 192 & +0.16     &  0.70     &  0.53     & --0.20    &  0.27     &  0.15     \\
NYAL &      &~24 & $\pm1.74$ & $\pm2.28$ & $\pm2.49$ &     & $\pm0.21$ & $\pm0.37$ & $\pm0.30$ & $\pm0.05$ & $\pm0.11$ & $\pm0.11$ \\
\hline
MATE & 1886 &~26 & --3.86    &  2.64     &  1.19     & 229 & --2.69    &  0.89     &  0.74     & --0.09    &  0.32     &  0.17     \\
ONSA &      &~29 & $\pm0.51$ & $\pm1.13$ & $\pm1.04$ &     & $\pm0.13$ & $\pm0.25$ & $\pm0.20$ & $\pm0.05$ & $\pm0.10$ & $\pm0.10$ \\
\hline
MATE & ~990 &~36 & --2.51    &  1.83     &  1.46     & 229 & --2.25    &  1.47     &  0.46     & --0.29    &  0.25     &  0.10     \\
WETT &      &~42 & $\pm0.40$ & $\pm0.82$ & $\pm0.71$ &     & $\pm0.10$ & $\pm0.19$ & $\pm0.17$ & $\pm0.03$ & $\pm0.05$ & $\pm0.05$ \\
\hline
MEDI & ~893 &~15 & --1.22    &  3.39     &  4.14     & 221 & --2.37    &  3.90     &  1.33     & --0.01    &  0.23     &  0.03     \\
NOTO &      &~19 & $\pm0.94$ & $\pm1.62$ & $\pm2.40$ &     & $\pm0.17$ & $\pm0.33$ & $\pm0.28$ & $\pm0.02$ & $\pm0.03$ & $\pm0.02$ \\
\hline
MEDI & 3776 &~28 & --0.55    &  3.13     &  3.31     & 201 & --0.73    &  1.61     &  1.15     & --0.28    &  0.41     &  0.08     \\
NYAL &      &~34 & $\pm1.13$ & $\pm1.87$ & $\pm1.92$ &     & $\pm0.23$ & $\pm0.38$ & $\pm0.34$ & $\pm0.06$ & $\pm0.11$ & $\pm0.11$ \\
\hline
MEDI & 1429 &~20 & --3.13    &  2.76     &  1.15     & 238 & --2.48    &  1.87     &  1.43     & --0.16    &  0.17     &  0.19     \\
ONSA &      &~25 & $\pm0.76$ & $\pm1.75$ & $\pm1.68$ &     & $\pm0.13$ & $\pm0.26$ & $\pm0.22$ & $\pm0.05$ & $\pm0.09$ & $\pm0.09$ \\
\hline
MEDI & ~522 &~20 & --2.14    &  2.56     &  1.25     & 238 & --2.41    &  1.55     &  0.88     & --0.31    &  0.06     &  0.10     \\
WETT &      &~22 & $\pm0.49$ & $\pm1.15$ & $\pm1.11$ &     & $\pm0.13$ & $\pm0.26$ & $\pm0.22$ & $\pm0.02$ & $\pm0.04$ & $\pm0.04$ \\
\hline
NOTO & 4580 &~23 & --1.62    &  6.02     &  3.73     & 189 & --0.05    &  0.39     &  0.44     & +0.02     &  0.36     &  0.24     \\
NYAL &      &~27 & $\pm1.30$ & $\pm2.34$ & $\pm2.32$ &     & $\pm0.22$ & $\pm0.31$ & $\pm0.32$ & $\pm0.09$ & $\pm0.11$ & $\pm0.10$ \\
\hline
NOTO & 2280 &~19 & --4.10    &  3.61     &  2.26     & 226 & --3.52    &  1.11     &  0.81     & --0.09    &  0.26     &  0.10     \\
ONSA &      &~24 & $\pm0.91$ & $\pm1.85$ & $\pm1.73$ &     & $\pm0.11$ & $\pm0.20$ & $\pm0.18$ & $\pm0.10$ & $\pm0.12$ & $\pm0.11$ \\
\hline
NOTO & 1371 &~21 & --3.39    &  2.51     &  1.66     & 226 & --2.98    &  1.81     &  0.92     & --0.21    &  0.21     &  0.12     \\
WETT &      &~22 & $\pm0.61$ & $\pm1.14$ & $\pm1.20$ &     & $\pm0.11$ & $\pm0.22$ & $\pm0.19$ & $\pm0.05$ & $\pm0.06$ & $\pm0.06$ \\
\hline
NYAL & 2387 &~31 & +2.17     &  4.03     &  1.56     & 206 & +1.44     &  1.28     &  0.50     & +0.22     &  0.63     &  0.25     \\
ONSA &      &~31 & $\pm0.86$ & $\pm1.47$ & $\pm1.52$ &     & $\pm0.15$ & $\pm0.25$ & $\pm0.22$ & $\pm0.06$ & $\pm0.12$ & $\pm0.12$ \\
\hline
NYAL & 3283 &162 & +1.67     &  2.49     &  2.06     & 206 & +1.03     &  1.41     &  0.24     & +0.24     &  0.42     &  0.10     \\
WETT &      &174 & $\pm0.25$ & $\pm0.49$ & $\pm0.52$ &     & $\pm0.14$ & $\pm0.22$ & $\pm0.21$ & $\pm0.06$ & $\pm0.12$ & $\pm0.12$ \\
\hline
ONSA & ~919 &~36 & --0.75    &  3.42     &  1.15     & 243 & --0.23    &  0.57     &  0.46     & +0.19     &  0.13     &  0.14     \\
WETT &      &~37 & $\pm0.49$ & $\pm0.93$ & $\pm0.96$ &     & $\pm0.08$ & $\pm0.13$ & $\pm0.13$ & $\pm0.04$ & $\pm0.07$ & $\pm0.07$ \\
\hline
\end{tabular}
\end{table}

It is also remarkable that influence of atmospheric loading on baseline
length rate is significant for many baselines.  Evidently, this effect
must be investigated more carefully and properly accounted during geodynamical
analysis.

\begin{figure}[ht]
\centering
\vskip 1em
\epsfxsize=140mm \epsffile{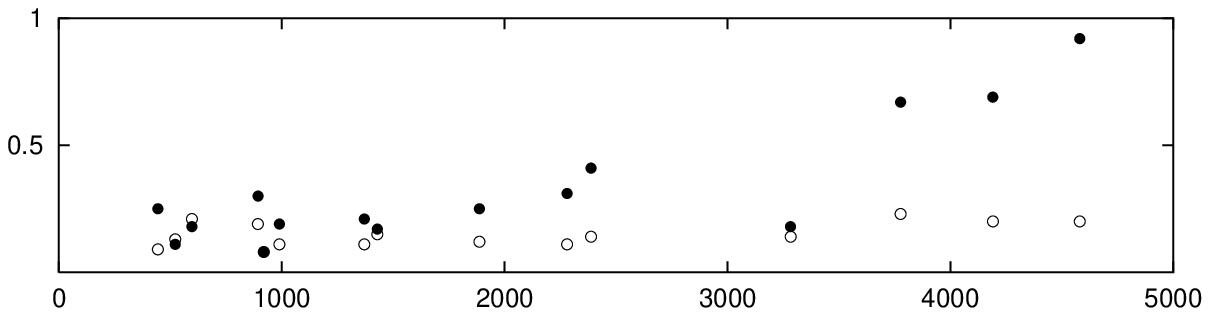}
\caption{Dependence of error in rate (mm) on baseline length (km) for
VLBI (filled circles) and GPS (light circles) data.}
\label{fig:lbacc}
\end{figure}

Figure \ref{fig:lbacc} shows dependence of error in baseline length rate on
length of baseline.  It is interesting that for GPS data error
is practically the same for all baselines unlike VLBI data.

Further steps of our work will include new re-computation of
EPN coordinate time series based on new combination of individual
EPN Analysis Center solutions, reprocessing of VLBI data with
new version of software, and more complete analysis of various
factors effected variations of baseline lengths. Analysis of variations
in vertical component of station displacement is also planned.

\section{Acknowledgement}

This research was partially supported by a grant of the
St.Petersburg Scientific Center of the Russian Academy of Sciences.

\end{document}